# Opto-UNet: Optimized UNet for Segmentation of Varicose Veins in Optical Coherence Tomography

Maryam Viqar
*Institute of Optical Materials and Technologies*
*Bulgarian Academy of Sciences*
Sofia, Bulgaria
maryamviqar92@gmail.com

Violeta Madjarova
*Institute of Optical Materials and Technologies*
*Bulgarian Academy of Sciences*
Sofia, Bulgaria
vdmadjarova@gmail.com

Vipul Baghel
*Department of Electronics Engineering,*
*Aligarh Muslim University*
Aligarh, India
vipulbaghel18elb127@gmail.com

Elena Stoykova
*Institute of Optical Materials and Technologies*
*Bulgarian Academy of Sciences*
Sofia, Bulgaria
elena.stoykova@gmail.com

*Abstract*— Human veins are important for carrying the blood from the body-parts to the heart. The improper functioning of the human veins may arise from several venous diseases. Varicose vein is one such disease wherein back flow of blood can occur, often resulting in increased venous pressure or restricted blood flow due to changes in the structure of vein. To examine the functional characteristics of the varicose vein, it is crucial to study the physical and bio mechanical properties of the vein. This work proposes a segmentation model Opto-UNet, for segmenting the venous wall structure. Optical Coherence Tomography system is used to acquire images of varicose vein. As the extracted vein is not uniform in shape, hence adequate method of segmentation is required to segment the venous wall. Opto-UNet model is based on the U-Net architecture wherein a new block is integrated into the architecture, employing atrous and separable convolution to extract spatially wide-range and separable features maps for attaining advanced performance. Furthermore, the depth wise separable convolution significantly reduces the complexity of the network by optimizing the number of parameters. The model achieves accuracy of 0.9830, sensitivity of 0.8425 and specificity of 0.9980 using 8.54 million number of parameters. These results indicate that model is highly adequate in segmenting the varicose vein wall without deteriorating the segmentation quality along with reduced complexity.

## I. INTRODUCTION

The extensive network of veins forms a vital part of human body as these vessels are responsible for maintaining the cardiac activity by carrying the blood. They are layered structures and physically they can be present in different ways like superficially underneath the fat layer, deep inside or like a connection between the deep and superficial ones. The veins have large diameter and thinner walls owing to the functional disparity between the veins and arteries. Commonly, veins experience force, pressure and stress. The variation in the forces can result in altered structure of veins leading to several diseases. Varicose vein is one such venous insufficiency disease present in the superficial veins, commonly seen in women. The dilation of veins, lead to back-flow of blood accompanied with alterations in vein pressure. This alters the physical components in the venous wall affecting the material properties [1]. Hence, it becomes crucial to elucidate the physical properties of the varicose vein for better disease diagnosis and progression related studies. One of most common findings is variations in the diameter of the veins in patients suffering from venous diseases [2]-[5]. In [6], reflux in veins has also been associated with the diameter of vessel. This makes segmentation of the venous walls crucial step in studying the physical alterations. In addition to this, while designing alternative vascular substitutes, segmentation is necessary to estimate the mechanical properties of vein [7]. Hence, segmentation is crucial to analyse the histological characteristics and biomechanical properties of vein.

In recent years, Optical Coherence Tomography (OCT) has been tremendously employed in ophthalmology settings. It a powerful 4D imaging modality that is based on low coherence interferometry. Light is used as the probing energy format and hence, it's a non-contact and radiation free technique. The attributes of 3D capture in OCT, allows the user to get the depth profile information which helps to extract enhanced information compared to confocal scanning microscopy related imaging-techniques. OCT is widely acknowledged as it provides scans having resolution of few micrometres paired with depth range of few millimetres [8]. These characteristics make OCT imaging a suitable technique for ex-vivo imaging of varicose vein to analyse structural variations with superior level of details in three-dimensions.

Segmentation of varicose veins acquired using the OCT can be manually done for each slice. However, this is a time-consuming task and requires a skilled professional. With the advent of artificially intelligent segmentation techniques for biomedical images, this task can be done using appropriate segmentation method. In this work, a novel enhanced architecture based on U-Net [9], is proposed for the segmentation of varicose vein layers.

The main contributions of this work can be summarised as follows:

- It is the first deep-learning based optimized network for varicose vein segmentation based on the symmetrical UNet architecture.

This project has received funding from the European Union's Horizon 2020 research and innovation programme under the Marie Skłodowska-Curie grant agreement No 956770.

- A new parallel-bridge block is proposed which employs atrous and separable convolution. The block is placed at the bridge of encoder-decoder structure of UNet which is responsible for reducing the model complexity and enhancing the segmentation accuracy.

The remaining section of the paper contains the following information: Section II illustrates the literature review of the relevant state-of-art methods, Section III describes the proposed modified model Opto-UNet in detail, Section IV gives the information about acquisition of the varicose vein dataset, experimental work and comparison of the results, Section V provides conclusion.

## II. Literature Review

Before the era of deep-learning, segmentation techniques were based on extraction of hand-crafted features like fuzzy convergence and illumination equalization [10], brightness variations and curvelet image analysis [11], energy-minimization with graph technique [12], gradient of pixels [13]-[15], etc. Indeed, the advent of deep-learning has paved way for more accurate and efficient methods in several domains of Computer Vision like classification, segmentation, denoising, etc. Semantic segmentation is one such area wherein labelling of pixel is done for segmentation problem. Utilising deep-learning models for semantic segmentation has been quite popular in recent years [16]-[23]. From biomedical imaging to road scenes, semantic segmentation has seen progressive growth. SegNet [16] was developed initially as a classification model consisting of 16 convolution layers with encoder-decoder type of structure. In order to label the pixels, the decoder uses the features generated by encoder along with skip layers. The model, quite often suffers due to missed or incorrect detection of pixels in the image due to lack of global information. PSPNet [17] is another segmentation model that performs pixel-level parsing in images using the residual connections and dilated network pyramid pooling algorithm with an enhanced capability to extract global level information. However, the method lacks low-level feature which makes it unsuitable for segmentation specially in bio-medical images where the region-of-interest usually occupy small areas and require high level of localization accuracy.

As far as biomedical image segmentation is considered, U-Net [9] stands out in comparison with several other methods. It is due to the fact that unlike other deep-learning based segmentation models, U-Net can be trained accurately with limited samples. The success for U-Net can also be attributed to deep layered encoder-decoder based architecture having skip-connections which allows the network to absorb high as well as low resolution features. U-Net stands strong in the domain of segmentation providing backbone to many architectures [20]-[23], specially designed for OCT images. However, network suffers due to the complex architecture utilizing millions of parameters which put constraints on the hardware. The skip connections in UNet only transfer information at the same level and lacks transfer of information to the subsequent layers which do not allow the model to absorb the multi-scale features. Furthermore, Residual Net [24] is a neural network where the sole idea of the model revolves around establishing identity mapped links between different layers of the network. to allow flow of information to the subsequent layers. The model is quite successful in resolving the issue of vanishing gradients and hence, helps to fight the problem of degradation of network specially during training in deeper networks. Another area of concern leading to inaccurate segmentation in UNet, is due to the direct upsampling of smaller receptive field areas obtained as a result of max-pooling on encoder side.

In this work, a model based on the architecture of UNet is proposed having residual connections to allow more feature interaction. A new block at bridge as shown in Fig.1 is proposed to extract segmentation specific enhanced features. The new block named as bridge-parallel block, uses dilated convolution [25] to capture wide area relationships between pixels. It performs sampling of the pixels whose sample rate is defined using the dilation rate. This ensures extraction of high-level structural information in images to address the accuracy loss encountered due to pooling and up-sampling operations. In addition to enhancing long-range spatial information, the proposed model also makes an attempt to employ disentangled depth information by using separable convolution which generates features in a depth wise manner followed by feature fusion. In summary, the proposed model follows a UNet shaped architecture having the encoder and decoder parts. The blocks in encoder and decoder structure have residual connections and at the bridge of the network the new proposed block performing atrous and separable convolution is employed.

## III. Methodology

A deep-learning model Opto-UNet is proposed here to segment the walls of the varicose vein as shown in Fig. 1. Elaborately, Opto-UNet means an optimized network based on U-net architecture designed for segmentation. The model borrows its backbone architecture from the versatile U-Net [9] whose structure is modified to attain enhanced performance both in terms of segmentation accuracy and model complexity. Furthermore, to improve the model's segmentation accuracy and computational efficiency, two more aspects are incorporated: (i) a parallel branch containing separable convolution and atrous convolution (ii) intermediate layers having residual connections.

The U-Net has an encoder-decoder structure performing the expanding and contracting operations to extract high- and low-level features. It also includes the skip connections between layers for transfer of features at different resolutions. Admirably, the information propagation between different levels of features extracted at varying resolutions opens room for highly accurate semantic segmentation in U-Net.

Though the deeper networks provide advantage of improved accuracy but they may also suffer with degradation due to higher training error rates. This commonly encountered problem of vanishing gradients can be addressed by stacking residual units in deeper networks. The residual connections [24] use identity mapping function which allows additional information flow through feature mapping from one layer to further deeper layers. The identity mapping connection used in this network is marked in Fig. 1 corresponding to block 1. In between the identity mapping connections, block has stacked layers of Batch Normalization (BN), activation function (ReLU) and convolution (conv) layers. Based on the performance outcomes presented in [26] for different combination of these layers, pre-activation unit with BN as first layer is implemented in this work.

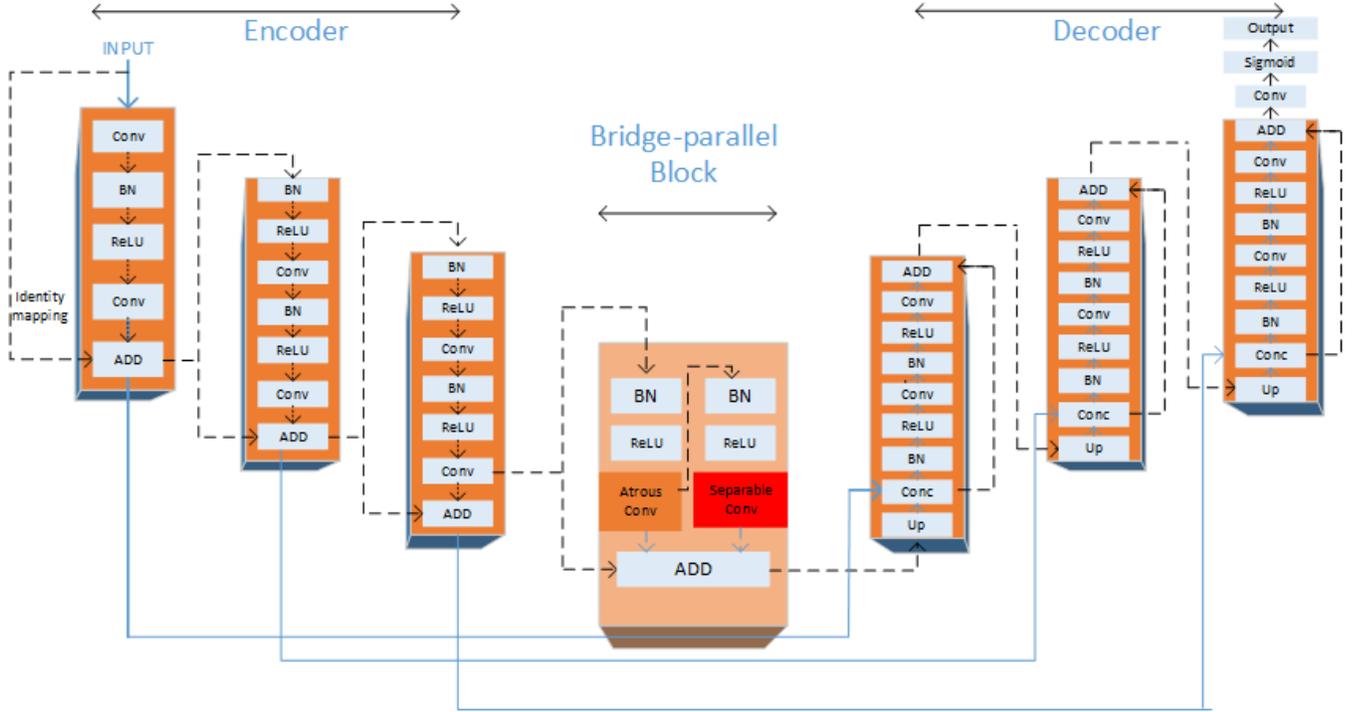

Fig.1 Block diagram of the proposed model Opto-UNet for varicose vein segmentation

The convolution used in U-Net simultaneously performs feature extraction and fusion. To disentangle the spatial and depth information the separable convolution [27] is used. The depth wise separable convolution performs splitting of channels along the depth to extract the feature first, then fuses them sequentially. Additionally, to capture the wide-range characteristics in images atrous convolution [25] can be used, which doesn't burden the network keeping the number of parameters same. But provide the advantage of improved segmentation accuracy by enabling the network to look at a bigger picture even at deeper layers. The atrous convolution can be defined as:

$$f[i] = \sum_n x[i + d.n]w[n] \qquad (1)$$

In (1), $f$ and $x$ are the output and input feature layers respectively, $i$ is the location on feature map, $w$ is the convolution kernel and $d$ is used to represent the dilation rate.

This work proposes a new block named bridge-parallel block inspired from the Optic-net [28] to inculcate atrous and separable convolution for optimized performance in form of parallel layers as demonstrated in Fig. 1. The block diagram shows that the output of these two parallel layers along with identity mapping function are merged together in the last layer using addition operation. The dilation rate is d=2 for both the branches performing atrous convolution to obtain a receptive field equal to 3×3 window. The feature maps are generated from two sources amongst which, one is spatial and other is spatial plus depth, which allows the network to learn varied characteristics simultaneously and hence, helps to optimize network's performance and complexity [28].

The complete architecture of the proposed model as portrayed in Fig.1 has different structure of building blocks on encoder and decoder side. The encoder unit has max-pooling layers along with fundamental unit comprising of BN, ReLU and regular convolution. The decoder side have the upsampling layer followed by concatenation (referred as 'conc' in Fig. 1) layer which receives information from the parallel block in encoder structure through the skip connection. The centre block referred as the bridge-parallel block has different structure from rest of blocks as defined above. This proposed unit is placed at the centre as it is last contraction stage where feature map is in the most reduced form containing important contextual information. After this, expansion using up-sampling (referred as 'Up' in Fig. 1) is done to enable precise localization. Owing to max-pooling in contracting part, loss of localization information is encountered. Introducing the bridge block at this stage compensates for the loss with its attributes of wider receptive field along with separate spatial and depth maps. As a result, this block helps to enhance the segmentation accuracy with reduced number of model parameters.

The aim of this network is to obtain segmented images from the input training set and the ground truth images for which loss function is required to evaluate weight parameters of the network. In this work, Dice Loss function obtained from Sørensen–Dice coefficient [29] is used for minimization of loss. Mathematically, it can be defined as follows:

$$DCL(R,B) = 2\frac{|R \cap B|}{|R|+|B|} \qquad (2)$$

$$DCL = \frac{\sum_{i=1}^{N} p_i y_i + \epsilon}{\sum_{i=1}^{N} p_i + r_i + \epsilon} + \frac{\sum_{i=1}^{N}(1-p_i)(1-y_i) + \epsilon}{\sum_{i=1}^{N} 2 - p_i - r_i + \epsilon} \qquad (3)$$

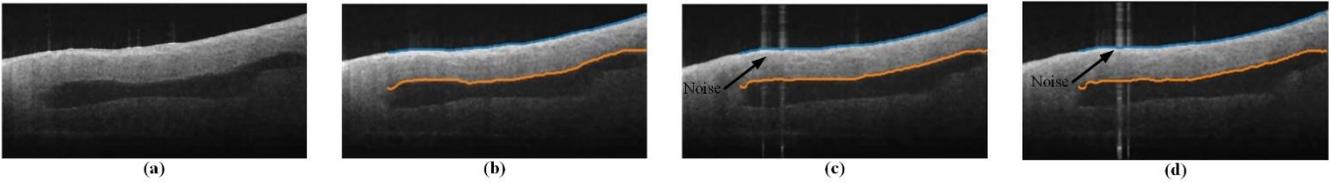

Fig. 2 (a) Original Varicose vein OCT image (b) Segmentation using Opto-UNet for Varicose vein (c), (d) Segmentation in presence of noise

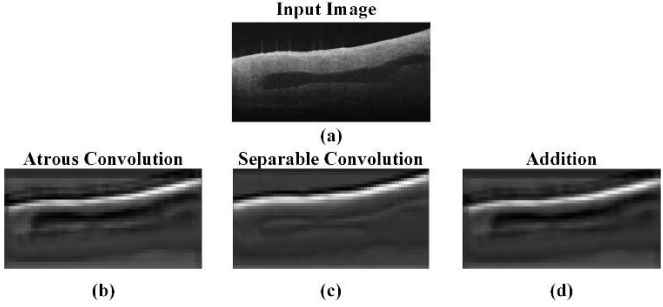

Fig. 3 (a) Original Varicose vein OCT image (b) Output for atrous layer of bridge-parallel block (c) Output for separable convolution layer of bridge-parallel block (d) Output for addition layer of bridge-parallel block

In (2), DCL represents the dice coefficient where R and B define the set of points in region of interest which is to be segmented i.e., foreground and the background region respectively. The image has total N pixels. Ground truth labels are represented using $y_i \in \{0,1\}$ while $p_i \in [0,1]$ represents the predicted labels in (3). To avoid zero valued denominator $\epsilon = 1$ is used. The loss is calculated by subtracting the coefficient from 1 as:

$$L = 1 - DCL \quad (4)$$

This loss function is used because of its inherent capability to solve the sample imbalance problem as our region of interest occupies lesser area compared to the background.

## IV. EXPERIMENTS

### A. Dataset Description

Varicose vein was extracted by surgical procedure at Medical Centre "Pirotska", Sofia, Bulgaria from a male patient, 57 years old suffering from the venous disease. Prior to the surgical procedure, informed consent was obtained from the patient. The veins after the surgical extraction were processed and stored as per the biological standard. This post-processing preparation work was done at the Institute of Mechanics, Bulgarian Academy of Sciences, Sofia, Bulgaria and images were acquired within 24 hours of the surgical procedure.

The varicose vein dataset was acquired on MHz Fourier domain OCT (Optores GmbH, Munich) system at Institute of Optical Materials and Technologies, Bulgarian Academy of Sciences, Sofia, Bulgaria. The OCT system uses Fourier Domain Mode Lock -FDML laser source at is 1.6 MHz sweeping frequency, 1309 nm central wavelength, and 100 nm sweeping bandwidth range. It provides a depth resolution of 17 μm. An area 7 mm x 7 mm (1024 by 1024 points) of the vein was scanned with lateral resolution 33 μm. The depth profile for each point in X-Y plane was generated by Fourier transform of the acquired signal that was rescaled to be linear in frequency space prior to the transform. The varicose vein dataset consists of 800 B-scans having resolution 512×256 pixels obtained after resizing due to the hardware limitations.

### B. Implementaion details

The proposed model was implemented on a NVIDIA GeForce RTX 3060 GPU with 12 GB graphic memory and 64 GB RAM. The network is trained and tested in Tensorflow 2.0 framework, with RMSprop's optimizer [30], learning rate of 0.0001, batch size 16 and dice loss [29] as the loss function. For training of the model, the dataset was divided into three sets namely training, validation and test containing 600, 100 and 100 images respectively.

### C. Results and Compariosn on Varicose Vein Dataset

In this section, we present the results obtained after the systematic evaluations of the proposed model for the varicose vein images in the acquired dataset. In order to make comparative analysis, we tested the other state-of-art deep-learning methods on the same dataset. The notable architecture used for comparison include Unet [9], SegNet [16], and PSPNet [17]. The models were trained in similar fashion and with the same environment as Opto-UNet for comparative evaluations.

The metrics used for evaluation of the proposed model and to compare with other main stream segmentation models are accuracy (ACC), sensitivity (SEN), specificity (SPE), dice similarity coefficient (DSC) and intersection over union (IOU). The equations of these metrics are as follows:

$$ACC = \frac{TP+TN}{TP+TN+FP+FN} \quad (5)$$

$$SEN = \frac{TP}{TP+FN} \quad (6)$$

$$SPE = \frac{TN}{TN+FP} \quad (7)$$

$$DSC = \frac{2*SEN*SPE}{SEN+SPE} \quad (8)$$

$$IOU = \frac{TP}{TP+FP+FN} \quad (9)$$

In (5)-(7), TP, TN, FP, FN are the number of pixels which are True Positive, True Negative, False Positive and False Negative in the output binary mask. Table 1 summaries the type of the model and its performance on the varicose vein dataset to make comparison of the Opto-UNet with other deep-learning models using the five metrics ACC, SEN, SPE, DSC and IOU. In addition to this, the number of parameters used in each of the models is also compared for the different models. As can be visualised from the table, the proposed models give the best results among all the models giving an accuracy of 0.9830 and number of parameters required is also the least, making it the suitable semantic segmentation model for varicose veins. The proposed bride-parallel block has significantly improved the model to achieve higher accuracy.

TABLE I. COMPARISION OF SEGMENTATION RESULTS FOR VARICOSE VEIN

| Model | ACC | SEN | SPE | DSC | IOU | PARAM-ETERS |
|---|---|---|---|---|---|---|
| UNet | 0.9791 | 0.8346 | 0.9968 | 0.9085 | 0.8210 | 31.04M |
| SegNet | 0.8868 | 0.1340 | 0.9031 | 0.2333 | 0.0245 | 29.45M |
| PSPNet | 0.9619 | 0.8142 | 0.9813 | 0.8899 | 0.7127 | 48.70M |
| Opto-UNet | **0.9830** | **0.8425** | **0.9980** | **0.9136** | **0.8395** | **8.54M** |

The performance of the SegNet model is highly inferior specially the SEN and IOU due to high false detections. The proposed model performs better than original UNet by a small margin in accuracy but nevertheless nearly 3.5-fold difference in number of parameters used, makes it highly efficient model. The segmentation results of the model can be seen in Fig. 2 with different type of images from the OCT varicose vein dataset. Fig. 2a presents the original OCT B-scan from the vein dataset; Fig. 2b shows the segmentation outcome after employing Opto-UNet. Moreover, to demonstrate the effectiveness of the segmentation method for images affected with noise, Fig. 2c and Fig. 2d are used wherein the presence of noise is marked by an arrow. It can be seen that even in presence of speckle noise, the method is capable enough to segment the layer with high precision. The results predicted by the proposed Opto-UNet model are almost perfectly segmented despite noise that can lead to artifacts otherwise. To visualise the significance of the operations performed by layers of parallel-bridge block namely (i) atrous convolution (ii) separable convolution and (iii) addition, the individual channel output for each layer is shown in Figs. 3b, 3c and 3d respectively. These figures illustrate the learning propagation of the input image through embedded layers of the parallel-bridge block of Opto-UNet model. The advantage of incorporating the atrous and the separable convolutions to extract spatial and depth level features can be observed with the procurement of fine-tuned feature map as the output of addition layer. Fig. 4 shows the changes in the loss and accuracy corresponding to the epochs during the training and validation phase. As the training epochs increase, the loss decreases and becomes quite stable nearly after 15 epochs. Similar results can also be visualised for accuracy curve. It can be inferred that the proposed model converges very fast and the learning rate is high. Hence, the proposed model is robust and accurate providing significant advantages for segmentation problems. However, there are limitations associated with the dataset as it contains images from only one patient suffering from varicose vein disease.

## V. CONCLUSION

Deep-learning networks are being used in several medical processing applications. In this work, we propose a model named Opto-UNet to address the segmentation of venous walls in varicose vein. Opto-UNet is extension of U-Net network which is an established segmentation model particularly for bio-medical images. A new block named bridge-parallel block is introduced for enhancing the performance using the concept of increased receptive field and generating depth-separable feature maps. The global and local level feature extraction of UNet architecture accompanied with residual connection, bridge-parallel block, has greatly enhanced the performance for segmenting the walls of the vein along with reduction in the amount of model parameters used. To perform comparisons, the acquired dataset was used to train and test the state-of-art deep-learning based segmentation models. The model gives quite promising results with faster convergence compared with other state-of-art models making it highly suitable for accurate segmentation of the varicose vein.

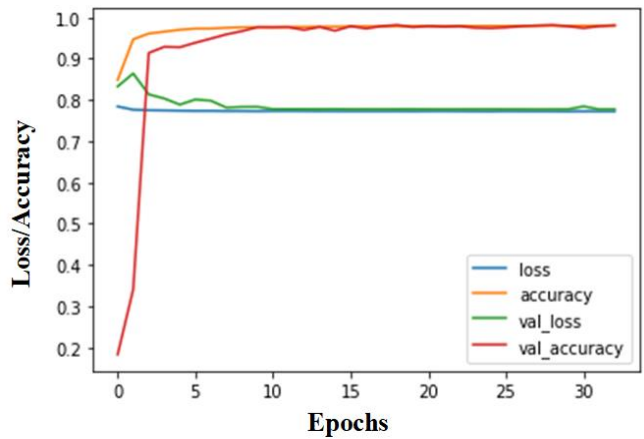

Fig. 4 Plots for training loss, validation loss, training accuracy and validation accuracy

In the future, more varicose vein data along with normal vein will be incorporated to validate segmentation results of the proposed Opto-UNet model. In addition, estimating physical properties like vein thickness, stress, etc. would be a fruitful area for research with wider applications.

ACKNOWLEDGMENT

M.V. would like to thank European Union's Horizon 2020 research and innovation programme under the Marie Skłodowska-Curie grant agreement No 956770 for the funding. V.M. thanks National Science Fund of Bulgaria (contract КП-06-Russia/7) and European Regional Development Fund within the Operational Programme "Science and Education for Smart Growth 2014–2020" under the Project CoE "National center of Mechatronics and Clean Technologies" BG05M2OP001-1.001-0008.